\documentclass[legalpaper,8pt]{jpconf}
\usepackage[]{fontenc}
\usepackage[latin9]{inputenc}
\usepackage{verbatim}
\usepackage{graphicx}
\PassOptionsToPackage{normalem}{ulem}
\usepackage{ulem}

\makeatletter

\pdfpageheight\paperheight
\pdfpagewidth\paperwidth

\newcommand{\noun}[1]{\textsc{#1}}
\providecommand{\tabularnewline}{\\}

\usepackage{graphicx}\usepackage{fontenc}
\newcommand{\maketitle}{}

\makeatother

\begin{document}
\textsf{\scriptsize{}Talk presented at the International Workshop
on Future Linear Colliders (LCWS15), Whistler, Canada, 2-6 November
2015}{\scriptsize \par}

\rule[0.5ex]{1\columnwidth}{1pt}

\title{Status of Dual-readout R\&D for a linear collider in T1015 Collaboration}

\author{C. Gatto$^{1}$, V. Di Benedetto$^{2}$ , E. Hahn$^{2}$and A. Mazzacane$^{2}$
\\
 \hspace{1.3cm}On behalf of T1015 Collaboration\cite{T1015}}

\maketitle
\address{$^1$ INFN, Sezione di Napoli, via Cinthia, 80126 Napoli, Italy}

\address{$^2$ Fermilab, Batavia (IL), USA} 

\ead{corrado.gatto@na.infn.it}
\begin{abstract}
The hadronic energy resolution required for an hadronic operating
at lepton collider  is at the limits or even exceeds that obtained
with traditional techniques. Furthermore, it is a well established
fact that the presence of an electromagnetic section in front of an
hadron calorimeter, as occurs in the layouts of the majority of detectors
operating at a collider, would deteriorate the hadronic energy resolution
of the device. The novel $ADRIANO$ technology (\textit{A Dual-readout
Integrally Active Non-segmented Option}), currently under development
at Fermilab, overcomes the above limitations by complementing an integrally
active calorimeter with the dual-readout technique. Detailed Monte
Carlo studies indicate that the energy resolution is in the $25\%/\sqrt{E}$
- $38\%/\sqrt{E}$ interval with a linear response of the detector
up to an energy of 200 GeV. A baseline configuration is chosen with
an estimated energy resolution of $\sigma(E)/E\approx30\%/\sqrt{E}$.
Several prototypes have been built by \textit{T1015} Collaboration
at Fermilab, to explore the effect of modifications of the layout
from the baseline configuration. Preliminary results from several
test beams at the \textit{Fermilab Test Beam Facility} (FTBF) of $\sim1\lambda_{I}$
prototypes are presented. Future prospects with ultra-heavy glass
are, also, summarized. 
\end{abstract}

\section{Introduction}

\label{sec:Introduction} The physics program at future lepton and
hadron colliders encompasses a very large number of processes involving
final states with multi-jets events. In such an environment, calorimeters
will play an important role at energies above 100 GeV, as their energy
resolution scales, in most cases, as $1/\sqrt{E}$ . An intensive
detector $R\&D$ and Monte Carlo simulation activity is already in
progress within the lepton colliders communities\cite{ILC-CLIC}.
\\
The general consensus within the lepton collider community is that
the jet energy resolution needed to successfully distinguish the W
from the Z signal in a high energy $(E_{cm}>500GeV$) is $\sigma(E)/E\approx30\%/\sqrt{E}$
or better. Such a resolution is unprecedented for conventional hadronic
calorimeters and it has been reached in the past only by massive compensating
calorimeters with very small volume ratio between passive and active
materials\cite{CHORUS}. Unfortunately the average density of detectors
whith such properties is relatively low. Consequently, the volume
needed to contain the showers would be so large to make their use
impractical in experiments with colliding beams. Furthermore, the
resolution of conventional, single-readout calorimeters is limited
by the fluctuations in the electromagnetic (\textit{EM}) content of
the hadronic shower and by the unequal response of such devices to
the \textit{EM} and hadronic components of the shower itself ($e/h\neq1$)\cite{WIGMANS}. 

In recent years, dual-readout calorimetry\cite{DREAM} has been introduced
as an alternative technique in order to cope with those effects. The
dual-readout technique relies on measuring event-by-event the \textit{EM}
fraction of the shower and is based on the simultaneous acquisition
of signals generated by independent shower production mechanisms,
thus providing complementary information on the composition of the
hadronic shower. 

Dual read-out calorimetry falls under two broad categories: sampling
and integrally active. Sampling dual-readout techniques are currently
investigated by the $DREAM$\cite{DREAM} Collaboration. Extensive
$R\&D$ conducted in recent years by $DREAM$ has confirmed the several
advantages of the dual-readout technique. However the sampling approach
introduces two new important sources of energy fluctuations: Poisson
fluctuations in the \u{C}erenkov signal, induced by the low photo-electron
statistics and sampling fluctuations, generated by the fact that the
absorber is totally passive and the signals are generated only in
a small fraction of the volume. Those fluctuations do not only impair
the energy resolution of hadronic particles and jets, but have also
detrimental consequences on the detection quality of photons and electrons.
The detection of \textit{EM} particles in high energy jets is similarly
affected. An obvious solution to the latter problem would be to design
a detector with two distinguished regions: a front \textit{EM} section
and a rear hadronic section. However, nowadays it is well understood
that such a configuration is sub-optimal in terms of energy measurement
of hadronic particles due to the extra fluctuations introduced by
the development of the shower across two different sections which,
in most cases, consist of media with very different properties~\cite{WIGMANS}.

Integrally active dual-readout techniques, on the other hand, are
mostly free of the above limitations since the the absorber is also
active and it participates in the compensation mechanism with the
production of the \u{C}erenkov signal. Test beams and extensive simulations
indicate that these techniques provide the energy resolution required
by the future lepton colliders and are capable of operate at the same
time as hadronic and \textit{EM} detectors, with no need for separate
devices. 

\smallskip{}

In this article we will give an update on the status of the Dual-Readout,
Integrally Active and Non-homogeneous Option ($ADRIANO$), based on
light signals produced in high transmittance optical glasses and scintillating
plastic materials. The related $R\&D$ is conducted as part of the
scientific program of Fermilab based, T1015 Collaboration. 

\section{Description of $ADRIANO$ techniques}

The $ADRIANO$ technique (\textit{A Dual-readout Integrally Active
Non-segmented Option}), developed by \textit{T1015}\cite{T1015} Collaboration
is intended to be used for High Energy as well as High Intensity experiements.
However, the different requirements for the two classes of experiments
have conducted to two separate designs, each optimized for that specific
application.

\emph{\uline{ADRIANO for High Energy experiements.}} Most of the
$ADRIANO$ prototypes which we have designed and built for High Energy
experiments have a full modular structure, with the base unit consisting
of an individual cell of parallelepiped shape with $40\times40\ mm^{2}$
cross-section and either $15\ cm$ or $25\ cm$ length. The cell consists
of a sandwich of scintillating fibers and high density, optical grade
heavy glass. The glass behaves as an absorber and as an active medium
at the same time, generating almost exclusively \u{C}erenkov light.
The scintillation and \u{C}erenkov sections of $ADRIANO$ are optically
separated. Therefore, the two generated light signals are well separated,
with minimal chance of cross-talk. We have considered various techniques
to optically separate those two regions: white and silver coating
of the glass, white coating, silver coating and aluminum sputtering
of the scintillating fibers and, finally, a thin layer of Teflon between
the glass plates and the array of scintillating fibers.

The scintillating fibers are either accommodated in grooves formed
in the glass itself or in white plastic trays sandwiched among plates
of glass. They run parallel to the longitudinal axis of the cell and
are responsible for the generation of the scintillation component
of the dual-readout calorimeter. The pitch between nearby fibers is
sufficiently small compared to the nuclear hadronic interaction length
of the detector that the shower sampling fluctuations are well contained.\\
 The \u{C}erenkov light generated inside the glass is collected by
WLS fibers running inside grooves parallel to the scintillating fibers
and optically coupled to the glass by compounds formulated ad-hoc
to match the large difference in the refractive indeces involved.
The two light components are read out at the back of each cell with
two distinct photodetectors. In some sense, $ADRIANO$ is a spaghetti
calorimeter with the passive absorber replaced by an active, transparent
absorber made of heavy glass. Another advantage of $ADRIANO$ relies
on the fact that the heavy glass absorber can be used to detect electromagnetic
showers in exactly the same ways as it has been done in the past with
lead glass based electromagnetic calorimeters. The latters are know
to be excelle \textit{EM} calorimeters in terms of detection afficiency
and energy resolution. Therefore, $ADRIANO$ does not require a front
electromagnetic section.

\begin{figure}[!]
\includegraphics[width=8cm]{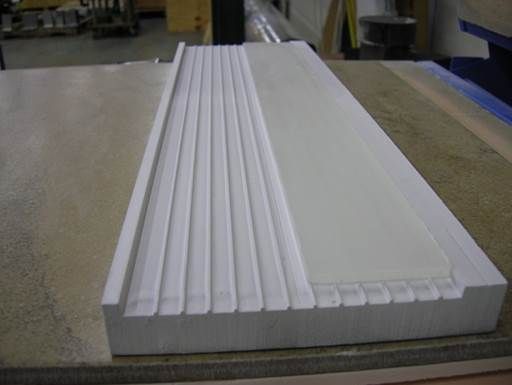} \includegraphics[width=8cm]{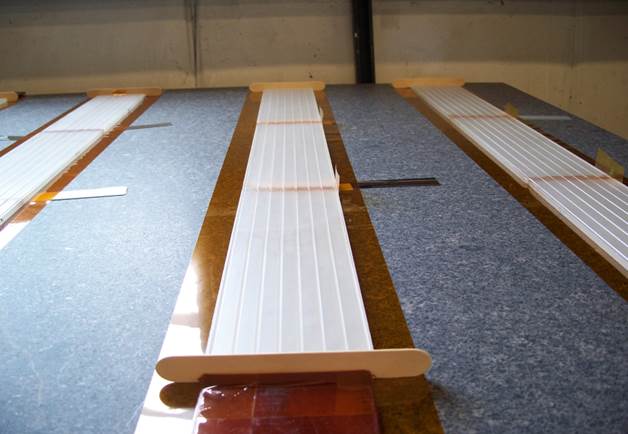}
\protect\caption{\textit{Molding fabrication technology}.}

\label{fig:mold} 
\end{figure}

Several heavy glasses (mostly lead and bismuth based) have been tested,
with the intent of comparing the \u{C}erenkov light yield and propagation.
Their refractive index ranges from $1.85$ through $2.24$ while the
densities range from $5.5g/cm^{3}$ through $7.5g/cm^{3}$. Various
constructions techniques have been considered: diamond machining,
precision molding, glass melting, laser drilling and photo-etching.
However, only the former two have been used for the production of
the eleven $ADRIANO$ prototypes presented in this report. A picture
of a $8\thinspace mm$ thick glass slice obtained with the precision
molding technique is shown in Fig.\ref{fig:mold} . 

Two new ADRIANO prototypes, each \emph{\noun{$10\,cm$}} wide and
$105\,cm$ long, have been built in 2014; both of them have been optimized
for experiments at High Energy lepton colliders. The first prototype
($ADRIANO\thinspace2014A$ ) was constructed by alternating 10 layers
of Schott SF57 lead glass plates, each 6.5 mm thick, with an equal
number of thin ($2\thinspace mm$) scintillating plates, extruded
at Fermilab's \textit{Plastic Extrusion Facility}. The scintillating
and \u{C}erenkov lights from the individual plates was captured with
optically coupled WLS fibers sitting in grooves running along the
detector. The second prototype ($ADRIANO\thinspace2014B$ ) was built
by stacking 10 layers of Schott SF57 lead glass plates, each 6.5 mm
thick. A total of 26 grooves were formed in the glass plates using
the precision molding technique. Twenty scintillating fibers and six
WLS fibers were accomodated in those grooves, the former (optically
separated from the glass) were responsible for the scintillating component
of the detector while the latter (optically couple to the glass) were
used to capture the \u{C}erenkov light. Both prototyped have been
completely built at the Fermilab's \textit{Thin Film Facility} and
were exposed to a beam of particles shortly after their construction.
Preliminary results of the measurements are presented below.

A detailed description of the layouts of the above prototypes and
their corresponding construction techniques will follow in an upcoming
article.

\begin{figure}[!]
\includegraphics[width=16cm]{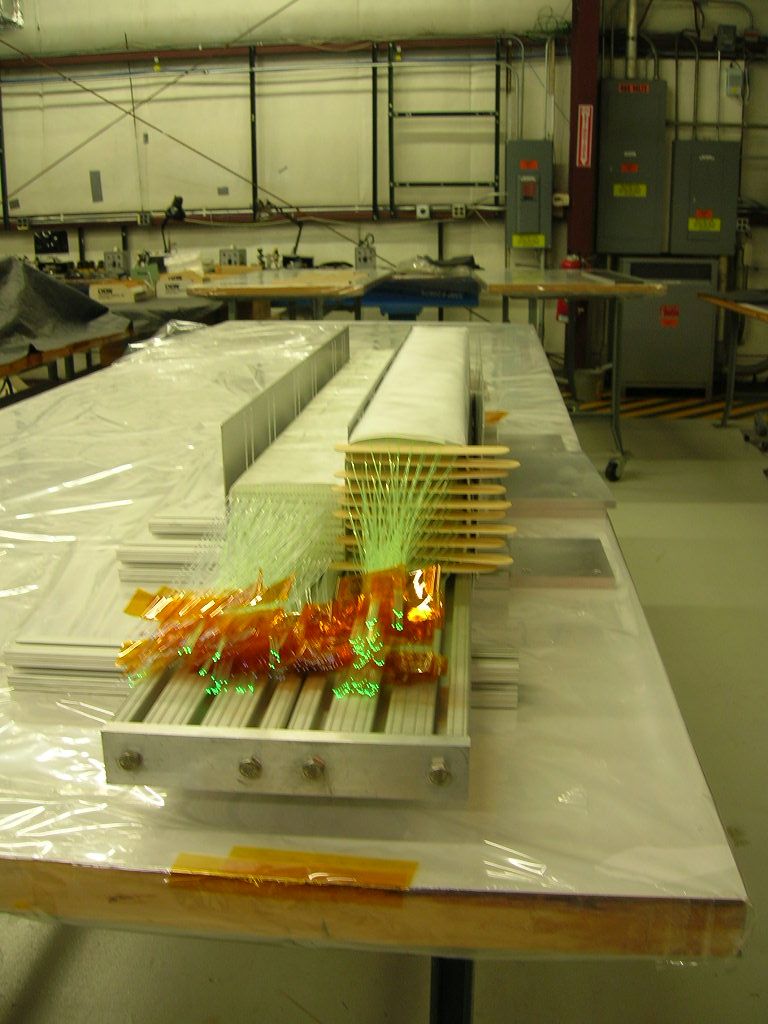} \protect\caption{\textit{ADRIANO2014A (right) and ADRIANO2014B (left). During the assembly
phase at Fermilab's Thin Film Facility.}}

\label{fig:ADRIANO2014} 
\end{figure}
A picture of $ADRIANO\thinspace2014A$ and $ADRIANO\thinspace2014B$
during the assembly phase is shown in Fig.\ref{fig:ADRIANO2014}.
\medskip{}

\begin{figure}[!]
\includegraphics[width=8.85cm]{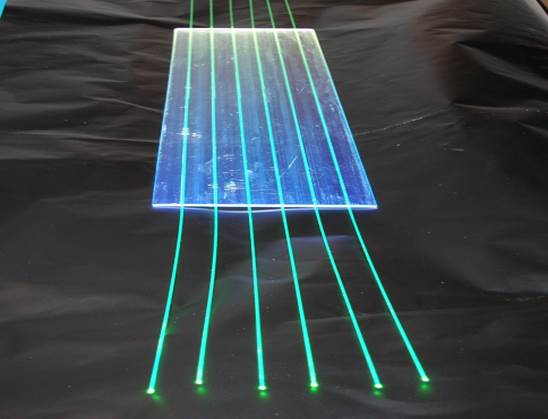} \includegraphics[width=8cm]{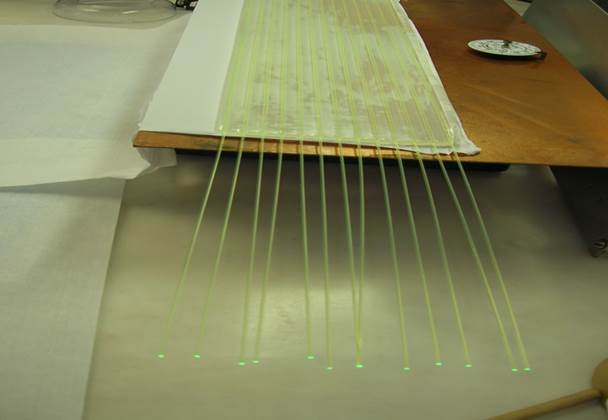}
\protect\caption{\textit{Plastic scintillator (left) and glass (right) plates in the
ADRIANO for High Intesity experiments}.}

\label{fig:mold-1} 
\end{figure}

\emph{\uline{ADRIANO for High Intensity.}}\emph{ }$ADRIANO$ prototypes
intended for High Intensity experiments are optimized for larger light
yield, rather than for high density, since they will operate mostly
at lower energies and in the \textit{EM} regime. They were, initially
designed for the ORKA and REDTOP projects at Fermilab, where they
would operate in the 5-500 $MeV$ energy regime, but the layout can
be properly optimized for any experiment with energy sensitivity above
few MeV. In this case, we replaced the scintillating fibers with plates
of thin ($2\,mm$) and grooved extruded scintillator sandwiched between
thicker (4.2 mm) heavy glass plates (Schott SF57), also grooved. Each
plate is 10 cm wide and 37 cm long and it is formed with the molding
technique described above. A picture of a the two plates is shown
in Fig.\ref{fig:mold-1} 

. 

The light readout of each plates uses WLS fibers: 6 for the plastic
plates and 13 for the glass plates. The larger number of WLS fibers
per unit surface of glass (0.031/cm$^{2}$ vs 0.012/cm$^{2}$ used
in the High Energy version) captures more \u{C}erenkov photons, at
the expenses of a lower average density of the detector and an increased
number of light sensors.

\section{ADRIANO readout system}

\label{sec:readout} The scintillating and WLS fibers from each $ADRIANO$
cell were bundled and routed each to a photodetector. In order to
compare the performance of the light collection system in various
situations we used three different photomultipliers (R647 and H3165
from Hamamatsu and P30CW5 from Sensetech) and two different SiPM ($4\times4mm^{2}$
square and $\O{2.7mm}$ round from FBK) for WLS fibers and only one
type of SiPM ($4\times4mm^{2}$ square from FBK) for the scintillating
fibers. When the PMT's were used, the fibers were routed through a
plastice fixture and coupled to the photosensor window with custom
made optical grease. In the case of SiPM's, we used either acrylic
light concentrators (designed and produced by INFN Trieste) in direct
contact with the fibers on one side and spaced $0.1m\mu$ from the
active SiPM surface or we routed the fibers directly to the SiPM up
to $0.1m\mu$ from the active SiPM surface. A picture of a INFN-Trieste
light concentrators is shown in top left of Fig.\ref{fig:testbeam2014}.
The output of the SiPM and PMT used was digitized by the TB4 DAQ system
developed at Fermilab. Among the features of the DAQ, the most relevant
for our application are:
\begin{itemize}
\item 50$\Omega$ inputs


\item 14 bit ADC;
\item $\sim30$ MHz bandwidth;
\item $\sim212$ MSPS digitizer;
\item Up to 16 channels per Motherboard;
\item Bipolar, so both positive (from SiPM) and negative (from PMT's) signals
can be acquired simultaneously;
\item Slow-control over USB, readout over 100 Mbit Ethernet.
\end{itemize}

\section{Results From the Test Beam at FTBF's }

\label{sec:Testbeam}

\begin{figure}[!]
\includegraphics[width=16cm]{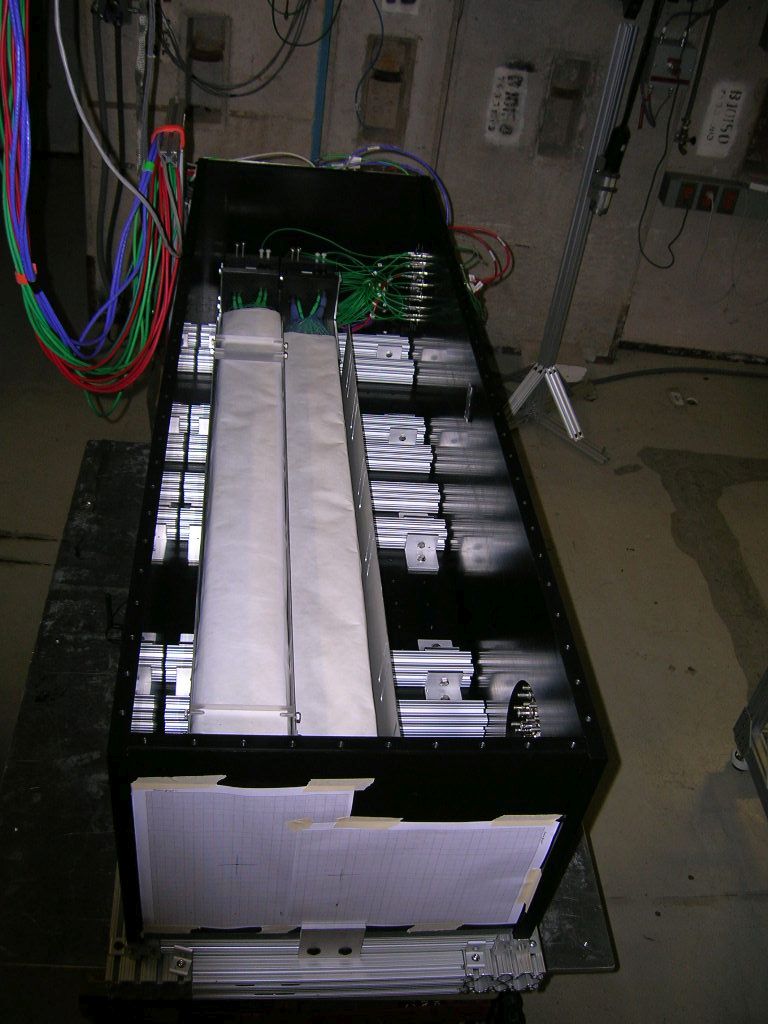} \protect\caption{Setup of a test beam at Fermilab's \emph{FTBF} of $ADRIANO\thinspace2014A$
and $ADRIANO\thinspace2014B$ prototypes (2014)}

\label{fig:testbeam2014} 
\end{figure}

The two prototypes for High Energy experiments: $ADRIANO\thinspace2014A$
and $ADRIANO\thinspace2014B$ were subject to several test beam during
the year 2014, at the \textit{FTBF Facility} of Fermilab (Batavia,
US). A picture of the test beam setup is shown in Fig.\ref{fig:testbeam2014}

\begin{figure}[!]
\includegraphics[width=8cm]{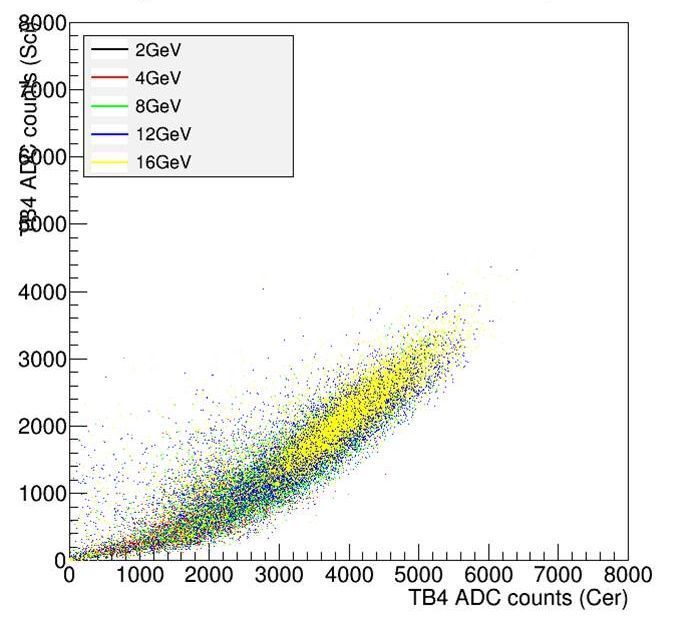} \includegraphics[width=8cm]{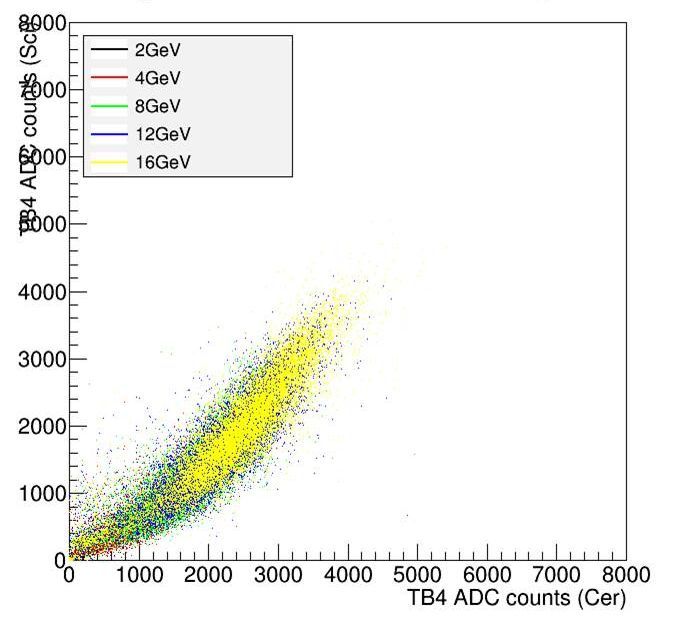}
\label{fig:SvsC} \protect\caption{\textit{Scatter plots of the scintillating vs the \u{C}erenkov light
for ADRIANO2014A (left) and ADRIANO 2014B (right)}.}
\end{figure}

The beam used was obtained by selecting secondary particles of known
momentum from a \emph{120 GeV} proton beam impinged on a tungsten
target. Data were taken at energies between 2 GeV and 1 GeV. The particle
identification was provided by FTBF's double-angle \u{C}erenkov systems.
The scatters plot for electrons with momentum varying between 2 GeV
and 16 GeV are shown in Fig.\ref{fig:SvsC} for $ADRIANO\thinspace2014A$
(left) and $ADRIANO\thinspace2014B$ (right). 

\begin{figure}[!]
\includegraphics[width=8cm]{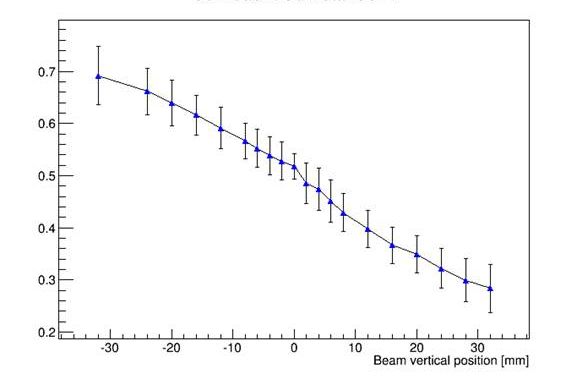} \includegraphics[width=8cm]{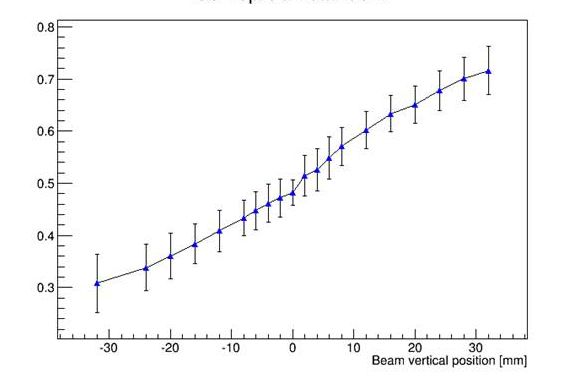}
\label{fig:verticalscan} \protect\caption{\textit{Scatter plots of the scintillating vs the \u{C}erenkov light
for ADRIANO2014A (left) and ADRIANO 2014B (right)}.}
\end{figure}

Analogously, the ratio between the \u{C}erenkov signal from the top
half and the bottom half of the detector obtained from a vertical
scan of a \emph{4 GeV} electron beam are shown in Fig.\ref{fig:verticalscan}
for $ADRIANO\thinspace2014A$ (left) and $ADRIANO\thinspace2014B$
(right). 

The error bars associated with the measurements indicate that the
Center of Gravity of a shower can be estimated with an error smaller
than 1 cm. Finally, the scintillation and the \u{C}erenkov light
yields have been estracted from the data analysis, by subtracting
the background induced by the pile-up of multiple events. The values
obtained for the two detectors tested are summarized in Table \ref{tab:Layouts-2014}. 

\begin{table}
\centering \resizebox{\textwidth}{!}{ %
\begin{tabular}{|c|c|c|c|}
\hline 
Prototype & Layout  & \u{C}erenkov L.Y. & Scintillation L.Y.\tabularnewline
\hline 
\hline 
2014A & 10 glass + 10 scintillating plates & 354 pe/GeV  & 523 pe/GeV \tabularnewline
\hline 
2014B & 10 glass plates + sparsified scintillating fibers & 338 pe/GeV  & 356 pe/GeV\tabularnewline
\hline 
\end{tabular}} \protect\caption{{\footnotesize{}Summary of light yields for }\textit{ADRIANO2014A
(left) and ADRIANO 2014B}{\footnotesize{} modules tested in 2014 at
Fermilab's }\emph{\footnotesize{}FTBF}{\footnotesize{}.}{\footnotesize \par}}

{\footnotesize{}\label{tab:Layouts-2014} }{\footnotesize \par}
\end{table}
The data indicate that the light yield is sufficient to garantee that
both techniques are capable of providing an hadronic energy resolution
of $\sigma(E)/E\approx30\%/\sqrt{E}$ in a detector volume which would
contain most of the shower.

\section{Summary}

\label{sec:Summary} Since the start of ADRIANO $R\&D$ program, we
were able to constantly improve the \u{C}erenkov light yield by refining
the construction techniques and the materials employed. A list of
the \u{C}erenkov light yield from the fifteen prototypes tested at
FTBF so far is summarized in TablecThe current limit for a $100\times90\times1050mm^{3}$
prototype for High Energy is roughly 360 pe/GeV, obtained at a test
beam at Fermilab in 2014 with the $ADRIANO\thinspace2014A$ prototype. 

\begin{table}
\centering \resizebox{\textwidth}{!}{ %
\begin{tabular}{|c|c|c|c|c|c|}
\hline 
Prototype \#  & Layout  & Glass  & g/$cm^{3}$  & L.Y  & Notes\tabularnewline
\hline 
\hline 
1  & 5 slices, machine grooved, unpolished, white  & Schott SF57HHT  & $5.6$  & $82$  & SiPM readout\tabularnewline
\hline 
2  & 5 slices, machine grooved, unpolished, white, v2  & Schott SF57HHT  & $5.6$ & $84$  & SiPM readout\tabularnewline
\hline 
3  & 5 slices, precision molded, unpolished, coated  & Schott SF57HHT  & $5.6$ & $55$  & 15 cm long\tabularnewline
\hline 
4  & 2 slices, ungrooved, unpolished  & Ohara BBH1  & $6.6$  & $65$  & Bismuth glass\tabularnewline
\hline 
5  & 5 slices, scifi silver coated, grooved, clear, unpolished  & Schott SF57HHT  & $5.6$ & $64$ & 15 cm long\tabularnewline
\hline 
6  & 5 slices, scifi silver coated, grooved, clear, unpolished  & Schott SF57HHT  & $5.6$ & $120$ & improved version\tabularnewline
\hline 
7  & 10 slices, white, ungrooved, polished  & Ohara PBH56  & 5.4  & $>30$  & DAQ problems\tabularnewline
\hline 
8  & 10 slices, white, ungrooved, polished  & Schott SF57HHT  & $5.6$ & $76$ & \tabularnewline
\hline 
9  & 5 slices, scifi Al sputter, grooved, clear, polished  & Schott SF57HHT  & $5.6$  & $30$ & 2 wls/groove\tabularnewline
\hline 
10  & 5 slices, ungrooved, polished  & Schott SF57HHT  & $5.6$ & $158$ & 2012 version\tabularnewline
\hline 
11  & 2 slices, plain  & Ohara experimental  & $7.5$ & $-$ & DAQ problems\tabularnewline
\hline 
12 & ADRIANO for ORKA - barrel (10 layers) & Schott SF57 & $5.6$ & $4000$ & Analysis in progress\tabularnewline
\hline 
13 & ADRIANO for ORKA - endcap (10 layers) & Schott SF57 & $5.6$ & $4000$ & Analysis in progress\tabularnewline
\hline 
14 & ADRIANO 2014A & Schott SF57 & $5.6$ & $354$ & 105 cm long prototype\tabularnewline
\hline 
15 & ADRIANO 2014B & Schott SF57 & $5.6$ & $338$ & 105 cm long prototype\tabularnewline
\hline 
\end{tabular}} \protect\caption{{\footnotesize{}Summary of the }\u{C}erenkov{\footnotesize{} light
yields for ADRIANO modules from seven test beams occurred from 2011
trough 2014.}{\footnotesize \par}}

{\footnotesize{}\label{tab:Layouts} }{\footnotesize \par}
\end{table}

\section{Future Prospects}

\label{sec:Future}

The first five years of $R\&D$ performed by the T1015 Collaboration
on the $ADRIANO$ technique have already produced clear directions.
Precision molding technique is, at present, the preferred fabrication
technique since it has the potential of making quick (less than 30
minutes) glass slices with optical surface finish and with appropriate
grooves. Nonetheless, we will keep exploiting other fabrication techniques
as they might have other potential advantages when compared to precision
molding. Photo-etching techniques, for example, are expected to be
equally fast and low-cost, although they have quite severe chemical
hazards associated with.

A great boost in $ADRIANO$ development is currently obtained with
Ohara sponsorship/partnership as they have provided bismuth glass
strips of commercial optical glass ($6.6g/cm3$, $nd=2.0$) and strips
of an even denser experimental glass with density of $7.54g/cm^{3}$
and refractive index of $2.24$.

Currently, one new $ADRIANO$ prototypes is under construction at
Fermilab aimed at High Energy experiments (high material density and
moderate light yied). The layout of the new prototype has been designed
based on the experience and the results obtained from the prototypes
built and tested in 2014. The goal is that of achieving an excellent
light yield, symilar to that of $ADRIANO\thinspace2014A$, and an
large average density, caracteristic of $ADRIANO\thinspace2014B$
prototype. A test beam for the new detector is already scheduled in
December 2015. For the medium future, we planned to continue to design
and build new modules and, consequently, to augment the volume of
the existing setup for a better shower containement.

\section{Conclusions}

\label{sec:Conclusions} In this report we have presented the status
of the R\&D activity performed by T1015 Collaboration on the $ADRIANO$
dual-readout technique. Two new detectors have been built and tested
in 2014 bringing the total number of prototypes to fifteen. Preliminary
results from a test beam at FTBF have been reported. The \u{C}erenkov
light yield we have obtained from most of the detector tested so far
is more than adequate for an hadronic calorimeter with an energy resolution
of $\sigma(E)/E\approx30\%/\sqrt{E}$ or better. Our goal is to increase
the light yield to a level where the \u{C}erenkov light yield of
$ADRIANO$ can be used to detect also electromagnetic particles. Hardware
$R\&D$ and Monte Carlo simulations are in an advanced state and constantly
providing directions in the design of new prototypes. Furthermore,
current results by T1015 obtained from several test beams have proved
that \u{C}erenkov light readout from heavy glasses with a WLS light
capturing technique is feasible and provides equal or better results
than traditional methods tipically employing large area PMT's directly
coupled to the glass. Correctly matching calorimetric techniques with
SiPM and Front End Electronics is also crucial for a good performance
of the detector. T1015 will address these issues in the future and
it will exploit new, glass-based, calorimetric techniques which will
also include scintillating heavy glasses.

\section{Acknowledgments}

\label{sec:Acknowledgments} We would like to thank Fermilab for its
on-going support and for providing all the infrastructure for the
construction and testing of $ADRIANO$ prototype. In particular, we
would like to thank Aria Soha, Todd Nebel, Geof Savage, Ewa Skupp
and Ray Safarik for their continuig assistence during the setup and
data taking of T1015 at FTBF. We would also like to thank Ohara for
providing us with free samples of their BBH-1 bismuth based glass
and of a new super-dense (about $7.5g/cm^{3}$) experimental glass
which have been used for the construction of ADRIANO prototype \#6
and \#11.\medskip{}

This work has been financially supported by the Fermilab's Detector
R\&D program and the ``TWICE-AGLAGE project'' of Istituto Nazionale
di Fisica Nucleare (Italy).

\section*{References}

\label{sec:References}


\providecommand{\newblock}{}
\begin{thebibliography}{}
\expandafter\ifx\csname url\endcsname\relax
  \def\url#1{{\tt #1}}\fi
\expandafter\ifx\csname urlprefix\endcsname\relax\def\urlprefix{URL }\fi
\providecommand{\eprint}[2][]{\url{#2}}

\end{thebibliography}


\begin{thebibliography}{1}
\bibitem{T1015} $www\-ppd.fnal.gov/FTBF/MOU_{P}DF/T1015\_mou.pdf$

\bibitem{ILC-CLIC} $http://www.linearcollider.org/$ $http://clic-study.web.cern.ch/clic-study/$

\bibitem{CHORUS} S. Buontempo et al., ``Construction and test of
calorimeter modules for the CHORUS experiment'', Nucl. Instr. and
Meth. A 349 (1994), p. 250.

\bibitem{WIGMANS} R. Wigmans, Calorimetry energy measurement in particle
physics, in: International Series of Monographs on Physics, vol. 107,
Oxford University Press, Oxford, 2000.

\bibitem{4TH-LOI} 4th Concept Collaboration - Letter of Intent from
the Fourth Detector (4th) Collaboration at the International Linear
Collider, 2009 also at: http://www.4thconcept.org/4LoI.pdf

\bibitem{DREAM-CRYSTALS} N.~Akchurin, et. al, ``New Crystals for
Dual-Readout Calorimetry'', Nucl. Instr. Meth. A604(2007)512-526

\bibitem{DREAM} All DREAM papers are accessible at http://www.\-phys.\-ttu.\-edu/dream,
and also http://high\-energy.\-phys.\-ttu.edu.particle physics,
in: International Series of Monographs on Physics, vol. 107, Oxford
University Press, Oxford, 2000. %

\bibitem{ILCROOT} $http://www.dmf.unisalento.it//~danieleb/IlcRoot/$\end{thebibliography}
\end{document}